\documentclass[%
prb,
nobibnotes,
superscriptaddress,
 amsmath,amssymb,
showpacs,
aps,
 reprint,%
]{revtex4-1}
\usepackage{hyperref}
\hypersetup{
colorlinks   = true, 
urlcolor     = blue, 
linkcolor    = blue, 
citecolor    = blue
}

\usepackage{graphicx}
\usepackage{dcolumn}
\usepackage{bm}
\usepackage{romannum}

\begin{document}

\preprint{AIP/123-QED}

\title[Vibrational signatures of diamondoid dimers with large intramolecular London dispersion interactions]{Vibrational signatures of diamondoid dimers with large intramolecular London dispersion interactions}

\author{Christoph Tyborski}
\affiliation{Institut f\"{u}r Festk\"{o}rperphysik, Technische Universit\"{a}t Berlin, Hardenbergstra\ss e 36, 10623 Berlin, Germany}
\author{Tobias H\"{u}ckstaedt}
\affiliation{Institut f\"{u}r Festk\"{o}rperphysik, Technische Universit\"{a}t Berlin, Hardenbergstra\ss e 36, 10623 Berlin, Germany}
\author{Roland Gillen}
\author{Tommy Otto}
\affiliation{Department Physik, Friedrich-Alexander-Universit\"{a}t Erlangen-N\"{u}rnberg, Staudtstra\ss e 7, 91058 Erlangen, Germany}
\author{Nataylia A. Fokina}
\affiliation{Institute of Organic Chemistry, Justus Liebig University, Heinrich-Buff-Ring 17 and Center for Materials (LaMa) Justus Liebig University, Heinrich-Buff-Ring 16, 35392 Giessen, Germany}
\author{Andrey A. Fokin}
\affiliation{Institute of Organic Chemistry, Justus Liebig University, Heinrich-Buff-Ring 17 and Center for Materials (LaMa) Justus Liebig University, Heinrich-Buff-Ring 16, 35392 Giessen, Germany}
\affiliation{Department of Organic Chemistry, Kiev Polytechnic Institute, pr. Pobedy 37, 03056 Kiev, Ukraine}
\author{Peter R. Schreiner}
\affiliation{Institute of Organic Chemistry, Justus Liebig University, Heinrich-Buff-Ring 17 and Center for Materials (LaMa) Justus Liebig University, Heinrich-Buff-Ring 16, 35392 Giessen, Germany}
\author{Janina Maultzsch}
\affiliation{Department Physik, Friedrich-Alexander-Universit\"{a}t Erlangen-N\"{u}rnberg, Staudtstra\ss e 7, 91058 Erlangen, Germany}


%

\date{\today}

\begin{abstract}
We analyze the vibrational properties of diamondoid compounds via Raman spectroscopy. The compounds are interconnected with carbon-carbon single bonds that exhibit exceptionally large bond lengths up to $1.71\,\mathring{\text{A}}$. Attractive dispersion interactions caused by well-aligned intramolecular H\raisebox{0.3mm}{$\cdots$}H contact surfaces determine the overall structures of the diamondoid derivatives. The strong van-der-Waals interactions alter the vibrational properties of the compounds in comparison to pristine diamondoids. Supported by dispersion-corrected density functional theory (DFT) computations, we analyze and explain their experimental Raman spectra with respect to unfunctionalized diamondoids. We find a new set of dispersion-induced vibrational modes comprising characteristic CH/CH$_{2}$ vibrations with exceptionally high energies. Further, we find structure-induced dimer modes that are indicative of the size of the dimers. 

\end{abstract}

\maketitle

\section{Introduction}
Diamondoids form a homologous series of well-defined, saturated hydrocarbons starting from the smallest possible unit of diamond, namely adamantane\cite{Landa1933,Dahl2003}. Higher homologs can be seen as assembled adamantane cages with various sizes and structures depending on how the cages are arranged\cite{Balaban1978,Dahl2003,Gund1974}. Being magnitudes smaller than bulk diamond (d$\leq\,$2nm), their mechanical rigidity or chemical inertness are comparable to those known in diamond\cite{Mckervey1980,Schwertfeger2008}. Optical properties, instead, are altered due to quantum confinement effects superimposed by structure-induced optical selection rules\cite{Landt2009,Richter2014,Banerjee2014}.\\ Recent developments in their chemical functionalization have led to new compounds with significantly altered physical properties\cite{Fokin2005,Gunawan2014,Fokina2007,Schreiner2006,Mella1996,Kahl2016}. For instance, the optical absorption energies can be tuned considerably by various functionalizations\cite{Rander2013,Landt2010,Banerjee2015,Voros2012,Zimmermann2013,Rander2017}. It was shown that the insertion of $sp^{2}$ defects as an interconnecting bridge between diamondoid monomers reduces the optical transition energies by $\sim2\,$eV\cite{Zhuk2015,Tyborski2017,Tyborski2017a,Meinke2014,Zimmermann2013,Banerjee2015,Meinke2013}. Such chemically blended diamondoids further exhibit vibrational properties characteristic for both the $sp^{2}$ and $sp^{3}$ diamondoid moieties\cite{Banerjee2015,Tyborski2017,Tyborski2017a,Tuinstra1970,Tyborski2015,Meinke2013}. The latter have been analyzed up to [12312]hexamantane both experimentally and theoretically\cite{Banerjee2014,Filik2006,Jensen2004,Jenkins1980}. It is, however, still an open question how certain functionalizations change their vibrational properties.\\ A new class of functionalized diamondoids was recently introduced, namely diamondoid dimers that are interconnected by very long carbon-carbon single bonds\cite{Schreiner2011,Fokin2012}. Depending on the size of the constituents, bond lengths up to $1.71\,\mathring{\text{A}}$ have been achieved\cite{Schreiner2011,Fokin2012}. The carbon-carbon bonds are stabilized by the intramolecular H\raisebox{0.3mm}{$\cdots$}H contact surfaces that act as dispersion energy donors\cite{Fokin2012,Wagner2015,Fokin2017}. On the one hand the dispersion interactions stabilize the interconnecting carbon-carbon bonds and cause thermal stabilities up to more than $200^{\circ}$C\cite{Schreiner2011,Fokin2012}. On the other hand they lead to high rotational barriers around the central carbon-carbon bonds up to 33\,kcal\,mol$^{-1}.$\cite{Fokin2012} It is therefore possible to access these compounds experimentally even at ambient conditions.\\ In this article we discuss the vibrational properties of such diamondoid dimers and related compounds. We especially address the influence of intramolecular dispersion interactions that lead to new characteristic vibrational modes. We support our analysis by dispersion corrected density functional theory (DFT) computations.

\section{Theoretical and experimental details}
\begin{figure}
\includegraphics[width=\columnwidth]{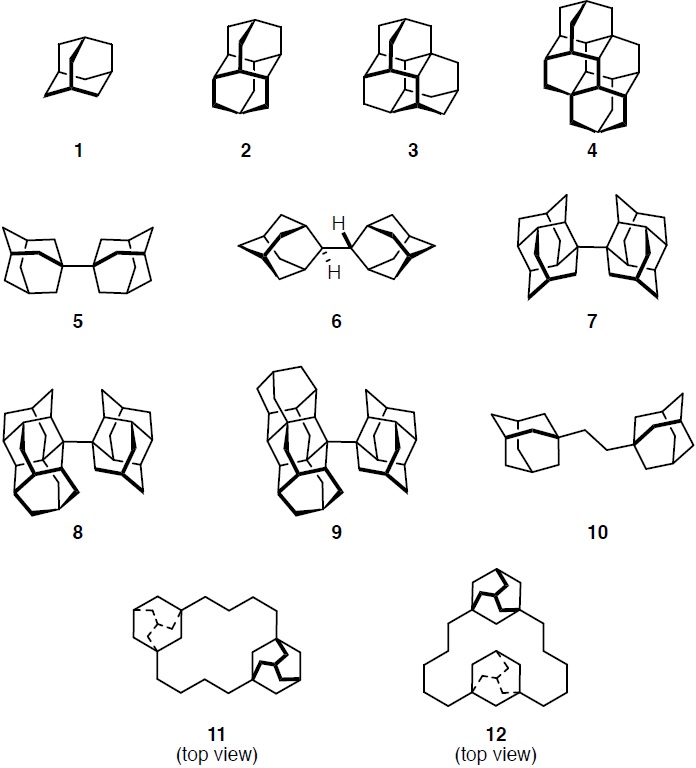}
\caption{Diamondoids (\textbf{1}-\textbf{4}), direct diamondoid dimers (\textbf{5}-\textbf{9}), and diamondoid dimers (\textbf{10}-\textbf{12}) labeled with compound numbers as used in the manuscript (Table \Romannum{1}).}
\label{Overview}
\end{figure}

Eight diamondoid dimers, built from adamantane \textbf{1} in Fig.~\ref{Overview}, diamantane \textbf{2}, triamantane \textbf{3}, and [121]tetramantane \textbf{4} were analyzed. They have either a single connecting carbon-carbon bond (compounds \textbf{5}-\textbf{9}) or carbon chains connecting the monomers (\textbf{10}-\textbf{12}). We will refer to compounds \textbf{5}-\textbf{9} as direct dimers in the following. Table~\Romannum{1} lists their chemical names and point groups.\\ All Raman spectra were taken in backscattering geometry with a Horiba LabRAM 800 spectrometer. We used a helium-neon laser providing a wavelength of 632.816\,nm ($\varepsilon_{\text{L}}=1.96\,$eV). Hence, all spectra are taken in non-resonant conditions, as the excitation energy is far below the optical transition energies of the dimers, which are expected to be higher than $\sim5\,$eV\cite{Zimmermann2013,Willey2005,Landt2009}. Excitation powers were always kept below 300\,$\mu$W to prevent sample degradation. At standard ambient conditions the analyzed diamondoid dimers form stable van-der-Waals crystals of which we used those facets that had the largest signal-to-noise ratios in the Raman spectra.\\ The computations of the vibrational eigenmodes of the diamondoid molecules were performed with the quantum chemistry code ORCA\cite{Neese2012}. Motivated by recent success on the accurate determination of atomic geometries and lattice vibrations in layered crystals\cite{Tyborski2017a,Gillen2018,Tornatzky2019}, we employed the PBE exchange-correlation functional\cite{Perdew1996} in combination with the Grimme-D3 semi-empirical dispersion correction\cite{Grimme2010}. 
The atoms were represented by all-electron def2-TZVP basis sets\cite{Weigend2005}. Using a tolerance of 10$^{-8}$\,eV for the SCF procedure and a Lebedev 434 points grid (grid5 in ORCA) to ensure sufficiently converged interatomic forces, we optimized the atomic positions until the residual interatomic forces were smaller than 0.01\,eV/\text{$\mathring{A}$} and subsequently computed the vibrational spectra and Raman intensities. The full width at half maximum (FWHM) in the modeled Raman spectra was set to 5\,cm$^{-1}$ for each Raman mode.

\section{Results and discussion}
\begin{table}
\caption{Names of each diamondoid compound and their point group as individual molecules.}
\begin{tabular}{ccc}
	\hline\hline
	No. & sample & point group\\\hline 
	\textbf{1} & Adamantane & T$_{d}$\\
	\textbf{2} & Diamantane & D$_{3d}$\\
	\textbf{3} & Triamantane & C$_{2v}$\\
	\textbf{4} & [121]Tetramantane & C$_{2h}$\\
	\textbf{5} & 1-(1-adamantyl)adamantane & D$_{3d}$\\
	\textbf{6} & 2-(2-adamantyl)adamantane & C$_{2}$\\
	\textbf{7} & 1-(1-diamantyl)diamantane & C$_{2}$\\
	\textbf{8} & 2-(1-diamantyl)triamantane & C$_{1}$\\
	\textbf{9} & 2-(1-diamantyl)[121]tetramantane & C$_{1}$\\
	\textbf{10}& 1,2-di(1-adamantyl)ethane & C$_{2h}$\\
	\textbf{11}& [4.4](1,3)adamantanophane & C$_{1}$\\
	\textbf{12}& [5.5](1,3)adamantanophane & C$_{1}$\\
	\hline
\end{tabular}
\end{table} The vibrational eigenmodes of diamondoids comprise characteristic carbon-carbon and carbon-hydrogen vibrations that are partly Raman active, depending on the symmetry of the monomer\cite{Filik2006,Jensen2004,Jenkins1980}. For instance, adamantane ($\text{C}_{10}\text{H}_{14}$) exhibits $T_{d}$ symmetry out of its 72 vibrational modes only 22 are Raman active\cite{Jenkins1980,Jensen2004,Landa1933}. However, due to very low intensities, only half of the allowed Raman modes are typically detected in an experiment\cite{Filik2006}. Larger diamondoids generally have more vibrational modes of which a larger fraction is Raman active since their symmetries gradually decrease with increasing sizes\cite{Dahl2003,Filik2006}.
\begin{figure*}
	\includegraphics{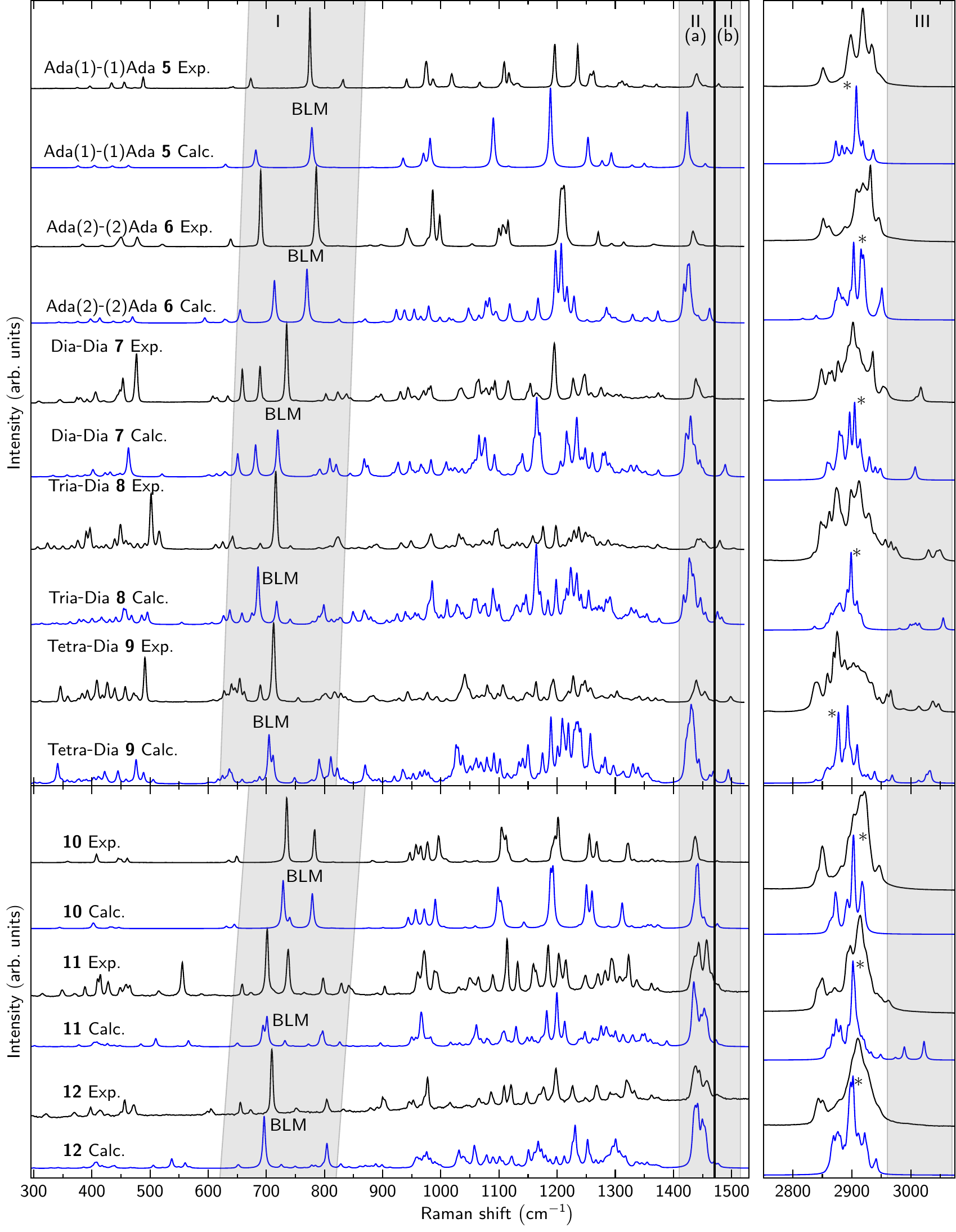}
	\caption{Experimental (black) and computed (blue) Raman spectra of single-bond diamondoid dimers are plotted. The laser excitation energy is $\varepsilon_{\text{L}}=1.96\,$eV. Ada, Dia, Tria, and Tetra stand for adamantane, diamantane, triamantane, and [121]tetramantane, respectively. Bold numbers correspond to the compounds as listed in Fig.~\ref{Overview}. Grey areas and Roman numbers highlight characteristic spectral regions in the spectra. Computed CH stretch frequencies ($\sim2950\,$cm$^{-1}$) are scaled by a factor of 0.976 and those with the largest intensities are labeled with asterisks. BLM stands for breathing-like mode. Computations were performed with the quantum chemistry code ORCA\cite{Neese2012} employing the PBE\cite{Perdew1996} exchange-correlation functional in combination with the Grimme-D3 dispersion correction\cite{Grimme2010}, see text.}
	\label{RamanI}
\end{figure*}
\begin{figure*}
	\includegraphics[width=2\columnwidth]{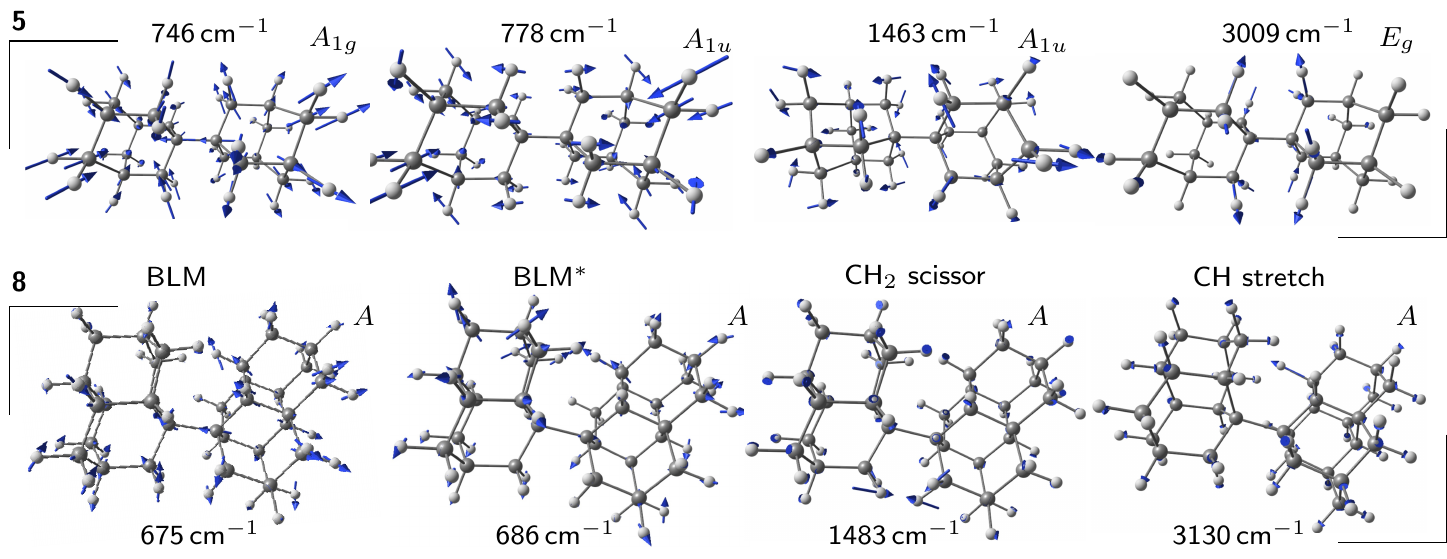}
	\caption{Characteristic eigenmodes of an adamantane-adamantane (upper row) and the triamantane-diamantane (lower row) dimer. The irreducible representations and computed frequencies are given next to the eigenmodes, respectively. BLM stands for breathing-like mode. In case of compound \textbf{5} ($D_{3d}$), the BLM splits into a symmetric ($A_{1g}$) and an antisymmetric ($A_{1u}$) eigenmode under inversion of which the latter is not Raman active. Two breathing-like modes are observed in compound \textbf{8} resembling those of pristine triamantane (lower frequency) and pristine diamantane (higher frequency). Both the eigenmodes for the CH$_{2}$ twist and CH stretch vibrations are those with the largest frequencies for the characteristic vibrations in the computations. Some displacement vectors are scaled for a better visibility. Computations were performed with the quantum chemistry code ORCA\cite{Neese2012} employing the PBE\cite{Perdew1996} exchange-correlation functional in combination with the Grimme-D3 dispersion correction\cite{Grimme2010}.}
	\label{EV1}
\end{figure*}
\begin{figure}
	\includegraphics{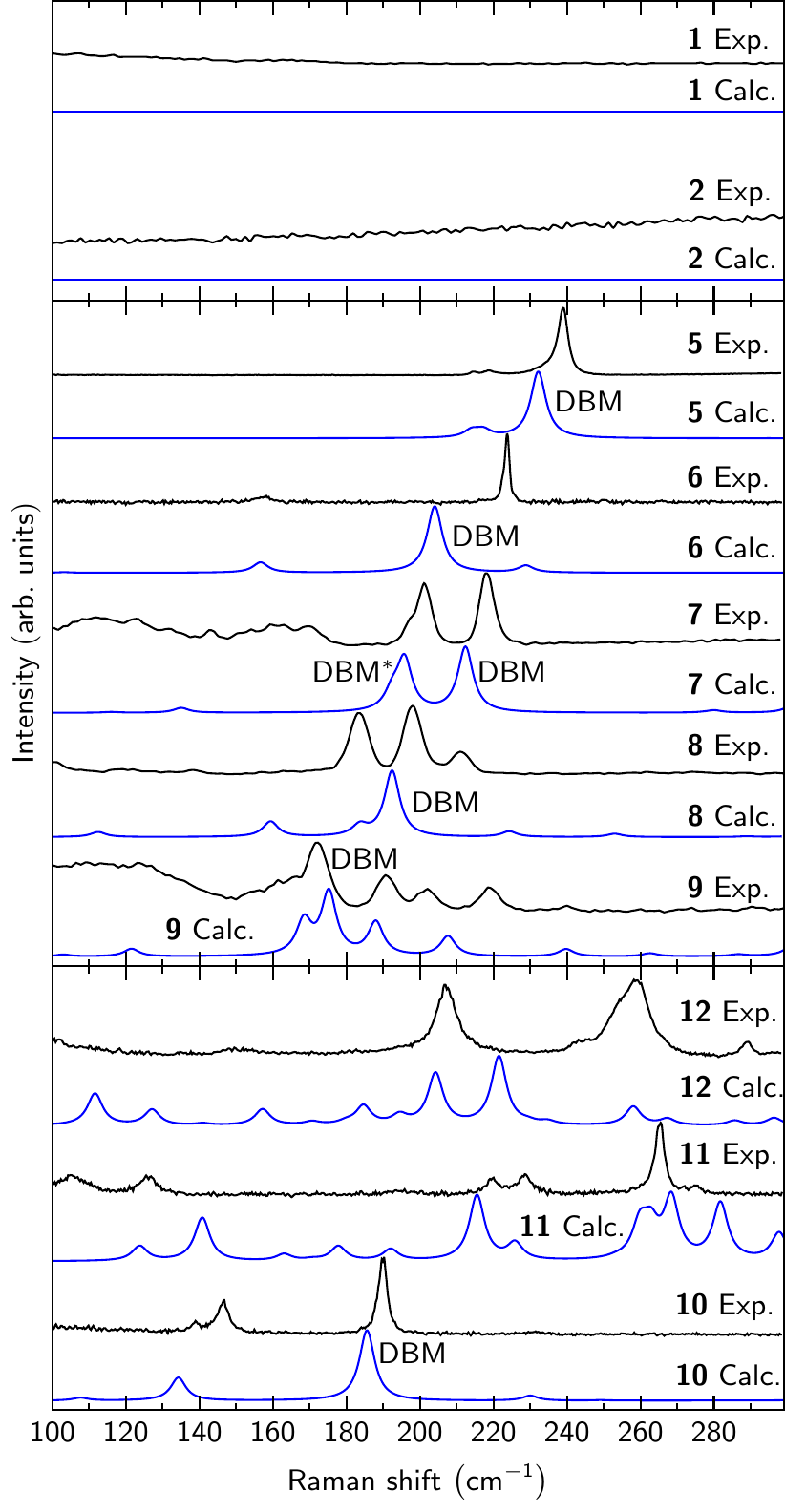}
	\caption{Experimental (black) and computed (blue) Raman spectra of different single-bond diamondoid dimers. The laser excitation energy is $\varepsilon_{\text{L}}=1.96\,$eV. The compounds are named as listed in Fig.~\ref{Overview}. DBM stands for dimer breathing mode\cite{Tyborski2017a}. For comparison, also the Raman spectra of pristine adamantane \textbf{1} and diamantane \textbf{2} are shown. Computations were performed with the quantum chemistry code ORCA\cite{Neese2012} employing the PBE\cite{Perdew1996} exchange-correlation functional in combination with the Grimme-D3 dispersion correction\cite{Grimme2010}.}
	\label{ULF}
\end{figure}
\begin{figure}
	\includegraphics{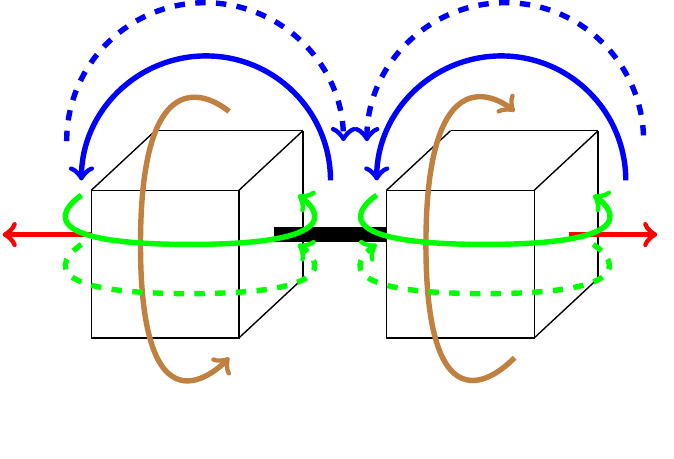}
	\caption{Schematic vibrational patterns of dimer modes in ascending order: rotational modes (brown), librations (blue dashed, green dashed), shear modes (blue, green), and dimer breathing modes (red).}
	\label{Scheme}
\end{figure}
\begin{figure*}
	\includegraphics[width=2\columnwidth]{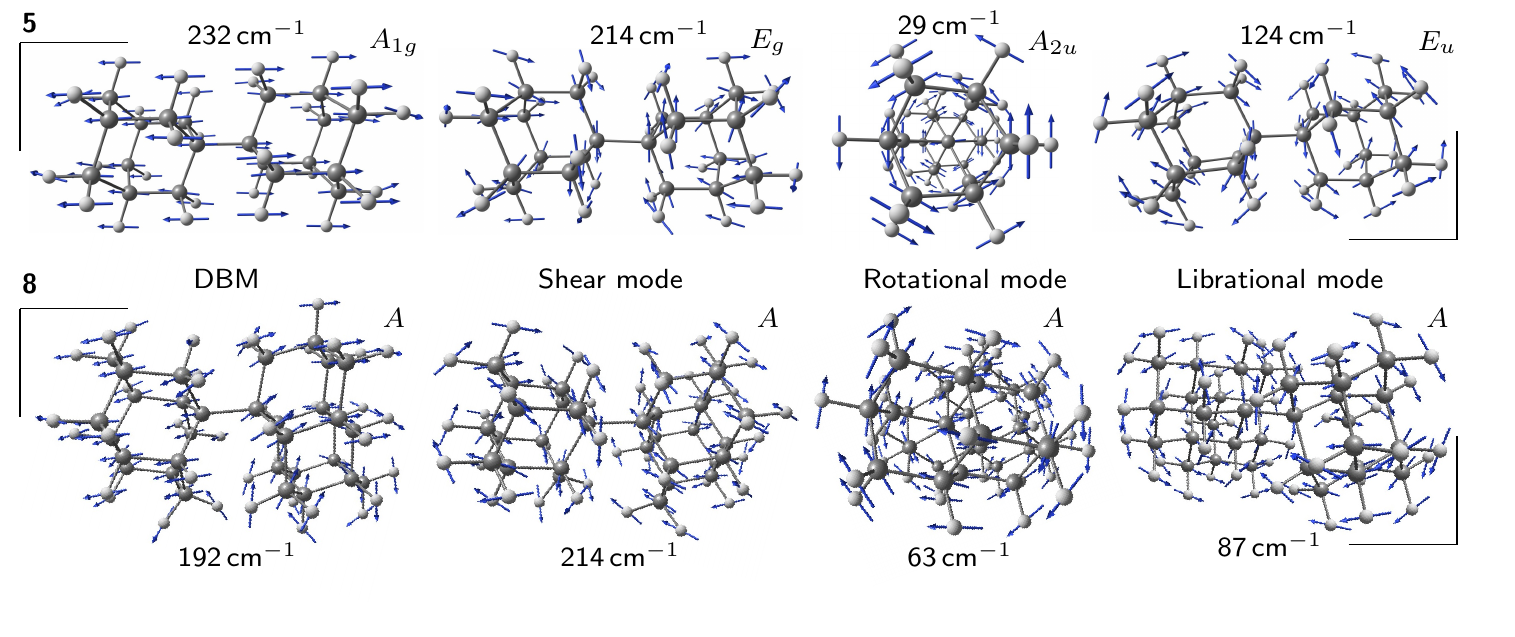}
		\caption{Characteristic low-frequency eigenmodes of an adamantane-adamantane \textbf{5} (upper row) and the triamantane-diamantane \textbf{8} (lower row) dimer. The irreducible representations and computed frequencies are given next to the eigenmodes, respectively. DBM stands for dimer breathing mode. Computations were performed with the quantum chemistry code ORCA\cite{Neese2012} employing the PBE\cite{Perdew1996} exchange-correlation functional in combination with the Grimme-D3 dispersion correction\cite{Grimme2010}.}
		\label{EV2}
\end{figure*}This, in particular, counts for the diamondoid dimers.\\ Figure~\ref{RamanI} shows experimental Raman spectra (black) in comparison to computed Raman spectra (blue) of all analyzed compound that are discussed in the following.\\ Up to $\sim1050\,$cm$^{-1}$ only carbon-carbon vibrations determine the Raman spectra in both pristine diamondoids and their single-bond dimers\cite{Jenkins1980,Jensen2004,Filik2006}. Lowest observable Raman modes with counterparts in unfunctionalized diamondoid monomers are CCC wagging modes. Yet, in adamantane ($T_{d}$) and diamantane ($D_{3d}$), Raman activity is suppressed due to their point-group symmetries. For example, in adamantane, the lowest-frequency vibrational mode is a three-fold degenerate $T_{1}$ wagging mode at around 350\,cm$^{-1}$.\cite{Jensen2004} Instead, in the single-bond dimers, their degeneracy is lifted. Due to very low intensities, however, these modes are barely observable both in the experimental and computed Raman spectra. We further observe an inertia-induced shift of the wagging modes down to $\sim250-330\,$cm$^{-1}$ in the dimers. The first low-frequency modes with reasonable Raman intensities are CCC bending modes. They have Raman active counterparts in the pristine diamondoids and range from $\sim350-650\,$cm$^{-1}$ partly exhibiting a mode mixing with other CCC wagging modes\cite{Jensen2004,Filik2006}. Again, we observe a downshift of their frequencies for an increasing size of the dimers, due to their larger inertiae. The same holds for the CCC stretching modes ranging from $\sim600-1050\,$cm$^{-1}$.\\ Especially between $\sim600-850\,$cm$^{-1}$, the size-induced downshift of Raman frequencies can be followed as indicated with the grey area \Romannum{1} in Fig.~\ref{RamanI}. This energy region contains the breathing-like mode (BLM) in unfunctionalized diamondoids, indicative for their sizes\cite{Meinke2013,Jenkins1980,Jensen2004,Filik2006}. In the dimers, the corresponding vibrations are still size-dependent, but further split into in-phase and out-of-phase vibrations as recently observed for analogous double-bond diamondoid oligomers\cite{Tyborski2017a}. According to the DFT computations, a significant elongation of the interconnecting bond is involved in the in-phase vibrations accompanied by a significant change of the polarizability. As a consequence, the in-phase vibrations exhibit a high Raman intensity, in contrast to the out-of-phase vibrations. In compound \textbf{5} ($D_{3d}$) the corresponding in-phase and out-of-phase modes are $A_{1g}$ and $A_{1u}$ modes, in compound \textbf{10} ($C_{2h}$), they are $A_{g}$ and $B_{g}$ modes, and in compounds \textbf{6}, \textbf{7} ($C_{2}$) they are $A$ and $B$ modes. A clear separation between in and out-of-phase vibrations in the remaining compounds (\textbf{8}, \textbf{9}, \textbf{11}, \textbf{12}) cannot be observed.\\ Especially in the homo dimers (\textbf{5}, \textbf{6}, \textbf{7}, \textbf{10}), the vibrational patterns of both BLMs are very similar to those of their unfunctionalized counterparts. In contrast, in the heterodimers, we rather observe joint breathing-like modes partly showing a mode mixing with other CCC stretching modes. The resulting BLM frequencies lie between those of the smaller and larger constituents. The larger decrease of BLM frequencies in compounds \textbf{11} and \textbf{12} compared to the other adamantane compounds (\textbf{5}, \textbf{6}) is due to a symmetric functionalization of the 1- and 3- positions in adamantane\cite{Jensen2004}. The twofold chemical alteration is accompanied by a larger perturbation of the primary eigenmode. This, in particular, results in a larger frequency reduction of the BLM derived vibrations\cite{Meinke2013,Filik2006,Jensen2004}. Characteristic eigenmodes of the BLM from two dimers can be seen in Fig.~\ref{EV1}.\\ Both in the computations and experimental Raman spectra, the BLM derived modes are the most intense ones in the spectral region of carbon-carbon vibrations (up to $\sim1050\,$cm$^{-1}$). They can therefore be well identified in the spectra (Fig.~\ref{RamanI}).\\ From $\sim1050\,$cm$^{-1}$ up to 3050\,cm$^{-1}$, the Raman spectra of the dimers comprise characteristic CH vibrations comparable to those in the unfunctionalized counterparts\cite{Jenkins1980,Jensen2004,Filik2006}. Their frequencies, however, are mostly not affected by the size or the structure of the dimers and are similar to those in, e.g., adamantane\cite{Jensen2004}. This can be seen, for instance, for the characteristic CH$_{2}$ scissoring modes at around $\sim1440\,$cm$^{-1}$, present in all pristine diamondoids and diamondoid dimers. They are denoted with \Romannum{2}\,(a) in Fig.~\ref{RamanI}. Besides, we observe new CH$_{2}$ scissoring modes that are only apparent in the direct dimers. Related frequencies are found around $\sim1490\,$cm$^{-1}$ both in the computations and experiments, \textit{i.e.}, at 50\,cm$^{-1}$ higher frequencies compared to unfunctionalized monomers. They are marked with \Romannum{2}\,(b) in Fig.~\ref{RamanI}. They generally exhibit vibrational patterns confined at the intramolecular contact surfaces. One example for the triamantane-diamantane dimer (\textbf{8}) is shown in Fig.~\ref{EV1}. It exhibits the largest computed frequency of a CH$_2$ scissoring mode in the particular compound. Repulsive interactions caused by the large H\raisebox{0.3mm}{$\cdots$}H inward contact areas lead to additional restoring forces that increase the frequencies of inward-oriented carbon-hydrogen vibrations.
 In case of the largest [121]tetramantane-diamantane dimer (\textbf{9}), we find three different scissoring modes at the intramolecular carbon cage edges with frequencies approx. 50\,cm$^{-1}$ larger than those at the outward dimer surfaces.\\ The smallest homo dimers (\textbf{5} and \textbf{6}), instead, do not show the high-frequency CH$_{2}$ scissoring modes neither in the experiments nor in the computations. Due to the compact structure of adamantane, the dimers do not form pronounced intramolecular surface areas. On the one hand, this absence leads to significantly shorter interconnecting carbon-carbon bonds than in the larger dimers (1.56\,$\mathring{\text{A}}$ in compound \textbf{6} vs. 1.71\,$\mathring{\text{A}}$ in compound \textbf{9})\cite{Fokin2012,Bond}. On the other hand, the dispersion-induced modification of CH$_{2}$ vibrations is significantly decreased. As a consequence, the CH$_{2}$ scissoring modes are rather delocalized and resemble those of the isolated monomers. This can be seen in Fig.~\ref{EV1} for the adamantane-adamantane dimer (\textbf{5}), where the CH$_{2}$ scissoring mode with the highest computed frequency is shown.\\The largest dispersion-induced frequency upshift can be observed for the inward-oriented CH stretch vibrations at around 3050\,cm$^{-1}$. Corresponding Raman modes are indicated in the shaded area \Romannum{3} in Fig.~\ref{RamanI}. Along with the symmetric (lower frequencies; $\sim2840-2890\,$cm$^{-1}$) and the anti-symmetric (higher frequencies; $\sim2890-2920\,$cm$^{-1}$) CH stretch vibrations, we find rather isolated, inward-oriented vibrations of opposite CH pairs belonging to either constituents (compare Fig.~\ref{EV1}). The more the hydrogen atoms are deflected towards the opposite diamondoid facet, the higher is the energy of the respective CH stretch vibrations. The newly observed vibrational modes range from $\sim2920-3050\,$cm$^{-1}$ in the larger diamondoid dimers. In contrast, none of the high-energy CH stretch modes can be observed in the adamantane dimers (compounds \textbf{5} and \textbf{6}), due to the reasons discussed before. The same holds for the other adamantane dimers interconnected with carbon chains (\textbf{10}-\textbf{12}). For the ring-like structures (\textbf{11} and \textbf{12}) we find several inward-oriented, CH stretch vibrations in the computations. However, only in the case of compound \textbf{11}, a structure-induced frequency increase of a few CH stretch modes is computed that cannot be observed in the experiments. We believe the mismatch is caused by slight differences in the adamantane orientation within the relaxed ground-state geometry, compared to the structure present in the van-der-Waals crystals\cite{Schreiner2011}.\\ In recently analyzed double-bond oligomers related high-frequency CH$_{2}$ scissoring and CH stretch modes were not observed\cite{Tyborski2017a}.
 The comparably stiff double bond causes a tilt of diamondoid moieties within the compounds impeding the formation of pronounced opposite intramolecular facets\cite{Zhuk2015, Meinke2014, Banerjee2015, Zimmermann2013}. Consequently, a structure-induced upshift of inward-oriented vibrations cannot be observed.\\ CH stretch vibrations with the largest intensities are totally symmetric, collective in-phase vibrations in the monomers ($A_{1}$ in compounds \textbf{1} and \textbf{3}, $A_{1g}$ in compounds \textbf{2} and \textbf{4})\cite{Jensen2004,Filik2006}. In the dimers they change to a $A_{1g}$ mode in compound \textbf{5} as well as $A$ modes in compounds \textbf{6}, \textbf{7}, and an $A_{g}$ mode in compound \textbf{10}. Due to the low symmetries (C$_{1}$), corresponding vibrations in the remaining compounds are all $A$ modes with vibrational patterns resembling those from the $A_{1}$ mode in adamantane (\textbf{11}, \textbf{12}) and the $A_{1g}$ mode in diamantane (\textbf{7}, \textbf{8}, \textbf{9})\cite{Jensen2004,Filik2006}. They are marked with asterisks in Fig.~\ref{RamanI}. We believe that the corresponding modes in the experimental spectra are also those ones with the largest intensities but they cannot convincingly be distinguished from other CH stretch modes. \\\\ We report a new set of structure-induced vibrational modes that we refer to dimer modes\cite{Tyborski2017a}. They comprise rotational modes, librations, shear modes, and dimer breathing modes as illustrated in Fig.~\ref{Scheme}, corresponding Raman spectra are shown in Fig.~\ref{ULF}.\cite{Tyborski2017a} Rotational modes are vibrations in which the whole monomers rotate with respect to each other around the interconnecting carbon-carbon bond. In the computations we find one rotational mode for each direct dimer, \textit{i.e.}, for compounds \textbf{5}-\textbf{9}. In case the symmetry is higher than $C_{1}$, the axis of rotation coincides with the principal axis of the respective compounds. We find frequencies between 29\,cm$^{-1}$ (\textbf{5}) and 77\,cm$^{-1}$ (\textbf{9}) in the computations. The smallest frequencies are found for the adamantane dimers, as the intramolecular surfaces and hence the intramolecular restoring forces for the rotational modes are only small. The next higher frequencies are librations, \textit{i.e.}, hindered rotations around the preferred intramolecular orientations. We find two librational modes for each dimer, which differ in their axes of rotation, with frequencies between $\sim90-135\,$cm$^{-1}$. For slightly higher frequencies, we find two shear modes along various directions for each compound. In contrast to the librational modes, shear modes are characterized by opposite displacements of the carbon atoms at the central connecting bond. The highest frequencies are found for the smallest dimers (\textbf{5} and \textbf{6}) up to 217\,cm$^{-1}$ whereas the largest dimer (\textbf{9}) exhibits frequencies only up to 175\,cm$^{-1}$. In contrast to the rotational modes and librations, the diamondoid carbon cages undergo significant deformation in each shear mode. Dimer modes with the largest frequencies are the dimer breathing modes (DBM). In double-bond interconnected diamondoid dimers, the DBM has an inverse dependence on the sizes of the compounds\cite{Tyborski2017a}. We can follow the same trend for the single-bond dimers both in the experiments and computations, as shown in Fig.~\ref{ULF}. Experimentally, frequencies between 170\,cm$^{-1}$ for the largest dimer (\textbf{9}) and 239\,cm$^{-1}$ for the smallest dimer (\textbf{5}) are observed.  Characteristic for the DBM is a large elongation of the central carbon-carbon bond and opposite deflections of the entire diamondoid moieties. Only in case of the high-symmetric adamantane dimer \textbf{5} ($D_{3d}$) the deflection directions of all carbon atoms coincide with the direction of the interconnecting carbon-carbon bond. In all other cases the position of interconnecting carbon-carbon bond does not agree with the diamondoid's centers of inertia. Their irreducible representations are $A_{1g}$ in compound \textbf{5}, $A_{g}$ in compound \textbf{10}, and $A$ in all remaining compounds. Vibrational patterns of all dimer modes are shown for two dimers in Fig.~\ref{EV2}. In contrast, the relevant region between $\sim100-300$\,cm$^{-1}$ in the Raman spectra of adamantane and diamantane (Fig.~\ref{ULF}) does not show any modes both in the computations and experiments, clearly indicating that the discussed modes are structure induced.\\ The ring-like compounds \textbf{11} and \textbf{12} do not exhibit a clear DBM or other dimer modes but rather show ring-like breathing modes\cite{Tyborski2017a}. In the low-frequency range we observe many different wagging modes from the interconnecting carbon chains resulting in various Raman allowed low-frequency modes for the compounds \textbf{10}-\textbf{12}. \\The DBM are the Raman modes with the largest intensities in the low-frequency region and can be well identified as shown in Fig.~\ref{ULF}. The diamantane dimer (\textbf{7}) shows two related DBM (DBM$^{*}$) in which diamantane moieties wag in slightly different directions. In the computations the Raman intensities are very close. Due to the overall good agreement to the experimental data, we believe both DBM/DBM$^{*}$ can be observed experimentally. Their large Raman intensities and size-dependent frequencies allow the identification of certain single-bond dimers. Especially in combination with an analysis of the BLM (indicative for the size of constituents) a clear assignment can be done. It even allows the differentiation between homo dimers that are interconnected at different positions. Both DBM and BLM of the adamantane dimers (\textbf{5} and \textbf{6}) are separated by $\sim20\,$cm$^{-1}$, although they contain the same diamondoid moieties. The single-bond dimers can be further clearly distinguished from $sp^{2}-$bound dimers, as the latter show the characteristic C\raisebox{0.25mm}{=}C vibrations at $\sim1660\,$cm$^{-1}$.\cite{Tyborski2017,Tyborski2017a,Banerjee2015} All remaining dimer modes have only very low Raman intensities both in the computations and experiments and can thus not be identified in the Raman spectra.\\\\ In conclusion, we have analyzed the vibrational properties of single-bond diamondoid dimers with exceptionally large dispersion interactions. The intramolecular van-der-Waals interactions cause a frequency upshift of isolated, inward-oriented CH$_{2}$ scissoring and CH stretch vibrations up to $150\,$cm$^{-1}$. Especially the high-frequency CH stretch modes at around 3050 can therefore be used as a direct marker of strong intramolecular H\raisebox{0.3mm}{$\cdots$}H interactions.\\We have shown a new set of structure-induced low-frequency modes comprising rotational modes, librations, shear modes, and dimer breathing modes (DBM) with frequencies between 23 and 232\,cm$^{-1}$. In all compounds, the dimer breathing mode exhibits the largest intensity. Its frequency depends inversely on the size of the compounds. In combination with the breathing-like mode (BLM), both size-dependent modes might be used as a fingerprint for the identification of single-bond diamondoid dimers.

  \section{Acknowledgments} The authors acknowledge financial support from the DFG under grant number MA 4079/6-2 and SCHR 597/24-1 within the Forschergruppe 1282 and Sonderforschungsbereich (SFB) 953 (B13). P.R.S. also acknowledges support from the DFG under grant number SCHR 597/27-2 (SPP 1807 "Dispersion"). The authors also gratefully acknowledge the "Regionales Rechenzentrum Erlangen" (RRZE) and the North-German Supercomputing Alliance (HLRN) for providing the computational resources.


\begin{thebibliography}{10}
\bibitem{Bond}
Bond lengths are computed with a dispersion corrected DFT approach as described
  in the theoretical details.
\end{thebibliography}
\end{document}